\documentclass[twocolumn,aps,prb,here,floatfix]{revtex4}
 
\usepackage{graphicx}
\usepackage{epsfig}

\begin{document}
\baselineskip11pt
\parindent0pt 

\title{Vortices and Edge Reconstruction in Small Quantal Systems at 
High Angular Momenta} 

\bigskip
 
\author{M. Toreblad, Y. Yu and S.M. Reimann}

\bigskip

\address{Mathematical Physics, Lund Institute of Technology, SE-22100 Lund, Sweden}

\author{M. Koskinen and M. Manninen}

\address{$^*$NanoScience Center, Department of Physics, FIN-40351 University of Jyv\"askyl\"a, Finland}
 
\begin{abstract}
Vortices can form when finite quantal systems are set to rotate.
In the limit of small particle numbers 
the vortex formation in a harmonically trapped {\it fermion} system, with repulsively interacting
particles, shows similarities to the corresponding boson system, with 
vortices entering the rotating cloud for increasing rotation. 
We show that for a larger number of fermions, 
$N\gtrsim15$, the fermion vortices compete and co-exist with 
(Chamon-Wen) edge-reconstructed ground states, forcing some ground states,
for instance the central single vortex, into the spectrum of excited states.
Experimentally, the fermion system could for instance be a semiconductor
heterostructure, a quantum dot, and the corresponding boson system a magneto
optical trap (MOT).
\end{abstract}

\maketitle

\section{Introduction}

Macroscopically, the appearance of vortices is an every-day 
phenomenon. Hardly anyone has missed the swirl of water going down the drain,
or the vortices forming while stirring in a cup of coffee.
Correspondingly, on the microscopic level the formation of 
vortices and vortex lattices is known to lower the energy of a rotating 
quantum system. A vortex is then characterised by a singularity
encircled by annular currents.

In the 50's Abrikosov predicted the existence of vortices in 
superconductors subject to a
magnetic field~\cite{Abrikosov}. Almost 50 years later, in 2001,
Abo-Shaer~\textit{et al.}~\cite{AboShaer} found
vortices forming in rotating Bose-Einstein condensates.
These many-body  
systems consist of paired electrons and bosonic atoms, respectively. 
Mean-field approaches are often employed to describe the complex
structures that these systems comprise.  
For Bose-Einstein condensates, one often applies the 
Gross-Pitaevskii equation. 
In this way, Butts and Rokhsar~\cite{butts} and later on Kavoulakis, Pethick
and Mottelson~\cite{KavMotPet,KavReiMot}
found successive transitions between stable patterns of 
singly-quantised vortices, as the angular momentum was increased. 

Surprisingly
vortex solutions, as discussed above, can also be found in small fermion 
systems with non-paired, repulsively interacting particles
at large angular momenta~\cite{saarikoski2004,toreblad}. 
Such systems could, for instance, be realized in quantum dots~\cite{rmp} 
(small semiconductor islands confining a finite number of electrons)  
at high magnetic fields. 
Incredibly enough, the vortex formation by repulsively interacting 
fermions at high angular momenta is very 
similar to the bosonic case~\cite{toreblad}: 
Irrespective of the system being bosonic or
fermion, vortex solutions govern the ground states 
in the limit of large angular momenta.

Setting a cloud of fermions rotating is equivalent to applying a 
strong magnetic field. For an electron in a quantum dot, 
at a certain field strength the system spin-polarises, and forms  
a compact droplet where the $N$ particles occupy neighbouring single-particle 
orbitals with angular momenta $m=0,1,2,....(N-1)$, all belonging to the 
so-called lowest Landau level (LLL) with radial quantum numbers $n_r =0$.   
This compact, so-called ``maximum density droplet'' 
(MDD)~\cite{macdonald} is the 
finite-size analog to integer filling factors in the bulk~\cite{prange}.
At larger fields or similarly, for higher total angular momenta, 
the droplet reconstructs its charge distribution.
Different scenarios have been suggested (see \cite{rmp} for a review):
While in the low-$N$ limit, the formation 
of holes at the centre was discussed by MacDonald, Yang and 
Johnson~\cite{macdonald,yang}, 
Chamon and Wen~\cite{chamon} suggested that for larger droplets, the 
energy could be lowered  by splitting off a (homogeneous) ring of electrons. 
In fact, it turned out later that the formation of a 
charge density wave along the edge was even more favourable~\cite{ReKoMa}.  

In this article we re-investigate the reconstruction 
of the maximum density droplet (MDD) in the light of vortex patterns
and the apparent analogy to the bosonic case. We show that,
in agreement with the earlier findings of Yang {\it et al.}\cite{yang}, 
reconstruction of the MDD begins from the edge rather than the 
dot centre if the electron number exceeds $N\approx15$. 
At smaller dot sizes, however, 
vortices enter the MDD from the edge, in much analogy 
to the bosonic case~\cite{toreblad}. 
The universality between fermion and boson
systems discussed by Toreblad~\textit{et al.}\cite{toreblad} holds up to
$N\approx15$. For higher $N$, not all vortex solutions 
appear as ground states. Instead edge reconstruction takes over
at the onset of the MDD break-up. In the limit of very large 
angular momenta, vortices can 
co-exist with the usual (Chamon-Wen) edge-reconstructed ground states.

\section{Many-body aspects of the harmonic oscillator} 

The Hamiltonian $H_{\Omega}$, which
we investigate, is of two-dimensional nature and describes
an interacting 
finite many-body system under rotation (or, equivalently, subject to a 
magnetic field, see appendix) which is confined by a harmonic oscillator 
potential: 
\begin{eqnarray} 
H_{\Omega}&=&\sum_{i=1}^{N} \left( \frac{\mathbf{p}^{2}_{i}}{2m} + \frac{1}{2} m
\omega_{0}^{2}\mathbf{r}^{2}_{i} \right) + 
\sum_{i <j}^{N} v({\bf r}_i -{\bf r}_j )-\Omega L_{z}=\nonumber\\
&=&H-\Omega L_z~. 
\label{hamiltonian}
\end{eqnarray}
Here $m$ is the particle mass and $\omega_{0}$ 
the confining frequency of the oscillator. 
This Hamiltonian has, in fact, become the standard choice to model 
finite quantal systems -- it equally well describes the 
physics of electrons trapped in a quantum dot~\cite{rmp}, as 
bosonic or fermion atoms in a magneto-optical trap~\cite{chris}.  
The two-dimensionality originates in the quantum dot from the 
semiconductor heterostructure. In the atom trap, it can be designed 
by the experimental set-up. In addition, strong rotational
motion lowers the effective confinement perpendicular 
to the axis of rotation. This leads to a ground state
with only the lowest energy level occupied 
along the $z$-direction \cite{ben,CooWilGun}. 

The operator $-\Omega L_{z}$ adds rotation at an angular
frequency $\Omega$. Since the total angular momentum, $L$, is a good
quantum number to $H_{\Omega }$, it effectively subtracts a
term $\Omega L$ from the energy of the Hamiltonian $H$. As a result a larger
$\Omega$ will create a
ground state with higher total angular momentum. Instead of seeking
solutions at varying $\Omega$, we follow the tradition in nuclear
physics and solve $H$ for a given $L$ to obtain the total energy,
$E_{L}$, and the many-body state $\vert \Psi_{F,B} (L) \rangle$. Note, that
the lowest state for a fixed $L$ is not necessarily the ground state
of $H_{\Omega }$. To find that state, we still need to subtract
$\Omega L$.

As is usually done for these systems, the interaction between the electrons
in a quantum dot is modelled by the ordinary Coulomb repulsion:
\[
v(\mathbf{r}_{i}-\mathbf{r}_{j})= 
 \frac{e^2}{4 \pi \epsilon \epsilon_{0}
  |\mathbf{r}_{i}-\mathbf{r}_{j}|}~.
\label{int}
\]
For trapped, dilute atomic gases
it is well known~\cite{chris} that the interaction between the particles
are more realistically described by a zero-range 
potential of the form $a \delta (\mathbf{r}_{i}-\mathbf{r}_{j})$,
where $a$ is the scattering length~\cite{chris}. 
However, we have recently shown that the dominating features in the
boson many-body spectra are to a large extent independent 
of the two-body interaction~\cite{toreblad}, and very similar results 
were obtained both for short-range and Coulomb two-body forces in the 
lowest Landau level. For ease of comparison with the electron system 
under consideration, we thus restrict this study to 
the usual Coulomb interaction, 
{\it both} in the fermion {\it and} boson case.
To simplify the comparison further, only
\emph{one spin state} will be considered: Spin-polarised
fermions and bosons with zero spin. 

In second quantisation, here for fermions, we write 
\begin{equation}
H  =  \sum_{i,j=1}^{\infty} \langle i \vert h \vert j \rangle
\hat{c}_{i}^{\dag} \hat{c}_{j}+\frac{1}{2} \sum_{i,j,k,l=1}^{\infty} \langle i
j \vert v_{coul}\vert k l \rangle \hat{c}_{i}^{\dag} \hat{c}_{j}^{\dag} 
\hat{c}_{l} \hat{c}_{k} ,  \label{secquant}
\end{equation}
where  $h$ stands for the single-particle part of the Hamiltonian
in Eq.~(\ref{hamiltonian}), 
$\left\{\mid  i \rangle \right\} _{i=1}^{\infty }$ is the harmonic 
oscillator basis and 
$\hat{c}_{i}^{\dag}$ and $\hat{c}_{i}$ are the corresponding
creation and annihilation operators.
One can now simply construct 
the full Hamiltonian matrix and diagonalise
it to obtain the energies and corresponding
many-particle states. 
The many-particle solution $\vert \Psi_{F} \rangle $ for fermions
is then given by an 
expansion in Slater determinants $\vert  \psi_{i} \rangle$, 
\begin{equation}\label{expansion}
 \vert \Psi_{F} \rangle = \sum_{i} c_{i} \vert  \psi_{i} \rangle.  
\end{equation}
For the bosonic, symmetric solution $|\Psi _B\rangle $, 
Eq.~\ref{secquant} is similar, with the fermion operators exchanged for the
corresponding boson operators $\hat{b}_i^{\dag}$ and $\hat{b}_i$. 

This so-called ``exact'' diagonalisation is the brute-force method of
many-body quantum mechanics. 
However, to name the method ``exact'' is misleading. A full
Hamiltonian matrix would be infinitely large, which of course would be
impossible to handle. 
A further complication is the fact that the size of the Hamiltonian matrix
rapidly increases, making calculations for more than about 
six particles extremely time-consuming. 
In view of the 
structure of the single particle levels for high magnetic
fields~\cite{rmp}, we thus restrict the single particle states 
to the above mentioned lowest Landau level (LLL). 
Physically this means that we ignore high
energy excitations which would mix the lowest and higher Landau
levels. For both bosons and fermions this is a common
restriction, which is justifiable for dilute quantum gases 
at very high angular momenta, or at very high magnetic
fields. It  allows us to extend the 
studies to systems with up to 25 particles.

\section{Appearance of vortices in bosonic and fermionic systems}

In weakly interacting bosonic systems, which are set rotating, 
successive transitions between
stable vortex states have been found~\cite{butts} for increasing angular
momentum. 
In particular, in a rotating and weakly interacting 
bosonic system with a short-range interaction, solutions of the 
Gross-Pitaevskii 
equation have revealed~\cite{KavMotPet,KavReiMot} a central one-vortex
solution for the ratio of angular momentum to particle number 
being equal to one, $l=  L_{1V}^{B}/N=1,$
a two-vortex solution at $l=L_{2V}^B/N\approx 1.75$
and a three-vortex state at
$l=L_{3V}^{B}/N \approx 2.1~$ (in atomic units).

Assuming similar vortex formation in the (unpaired) fermion system the
vortex angular momenta are shifted by $N(N-1)/2$ as
compared to the boson system.
This shift corresponds to the many-body
configuration of the MDD, which is the 
fermion equivalent to the boson many-particle ground state. 
An intuitive explanation is given by the configurations of the Fock states 
in the LLL, i.e. $\mid n_0 n_1 n_2 \cdots n_m\cdots 
\rangle $, where $n_m$ is the 
number of particles in the single-particle state with angular momentum 
$m$. 
In the non-rotating ground state in the bosonic case 
all $N$ spin-less particles are occupying the 
lowest single-particle orbital, the 
Fock state being of the form $|N0000\cdots 000\rangle $. For spin-polarised 
fermions, the Pauli principle demands single occupancies up to the 
Fermi surface, and the Fock state of the MDD
is a single Slater determinant of the form 
$|111\cdots 1111000\cdots000\rangle $.
As the angular momentum of the MDD equals 
$N(N-1)/2$, a central single vortex in the fermion case should appear at
 $l= (L_{1V}^{F}- L_{MDD})/N=1$,
a two-vortex solution at
$  l=(L_{2V}^{F} - L_{MDD})/N \approx 1.75$, 
and a three-vortex solution correspondingly  at
$l=(L_{3V}^{F} - L_{MDD})/N \approx 2.1$.

\section{Comparison between rotating fermion and boson
    system}\label{comrot}

Let us now investigate the rotating fermion system as compared to the
rotating boson system. 
In the lowest Landau level, at fixed angular momentum  
$L$ the total energy is given by 
\begin{displaymath}
E_{L}=\hbar \omega_{0} ( L + N ) + \langle \Psi_{F,B}(L) \vert v \vert
\Psi_{F,B}(L) \rangle, 
\end{displaymath}
where the first term originates from the kinetic energy and the
confining potential, and the second term  from the interaction
energy. An increase in $\hbar \omega_{0} (L+N)$ indicates a centre of
mass excitation, whereas variations in $\langle \Psi_{F,B}(L) \vert v \vert
\Psi_{F,B}(L) \rangle$ imply changes
in the interaction energy, which affects the configuration.\\
Computationally, there now exist two more simplifying circumstances: (1)
The kinetic energy and the confining potential constitute a
diagonal term to the total energy. It is therefore sufficient to diagonalise only $
\langle \Psi_{F,B}(L) \vert v \vert 
\Psi_{F,B}(L) \rangle$ and add the term $\hbar \omega_{0} ( L + N )$ at the
end to obtain the total energy. The outcome will be the same as
including the diagonal terms in the diagonalisation. (2) The interaction is independent of
confining frequency, $\omega_{0}$, since  
\begin{displaymath}
\langle \Psi_{F,B}(L) \vert v \vert
\Psi_{F,B}(L) \rangle \vert_{\omega_{0}=\omega} =\sqrt{\omega} \langle \Psi_{F,B}(L) \vert v \vert
\Psi_{F,B}(L) \rangle 
\vert_{\omega_{0}=1}.
\end{displaymath}
This can easily be verified by writing the expression with second
quantisation and comparing the integrals for $\omega_{0} = \omega$ and
$\omega_{0}= 1$ a.u. The results presented are thus independent of confining frequency and it is
only necessary to study $\omega_{0}=1$ a.u. as long as only the
lowest Landau level is considered.
\begin{figure}[!h]
\unitlength1cm
\centerline{\epsfxsize=9cm\epsfbox{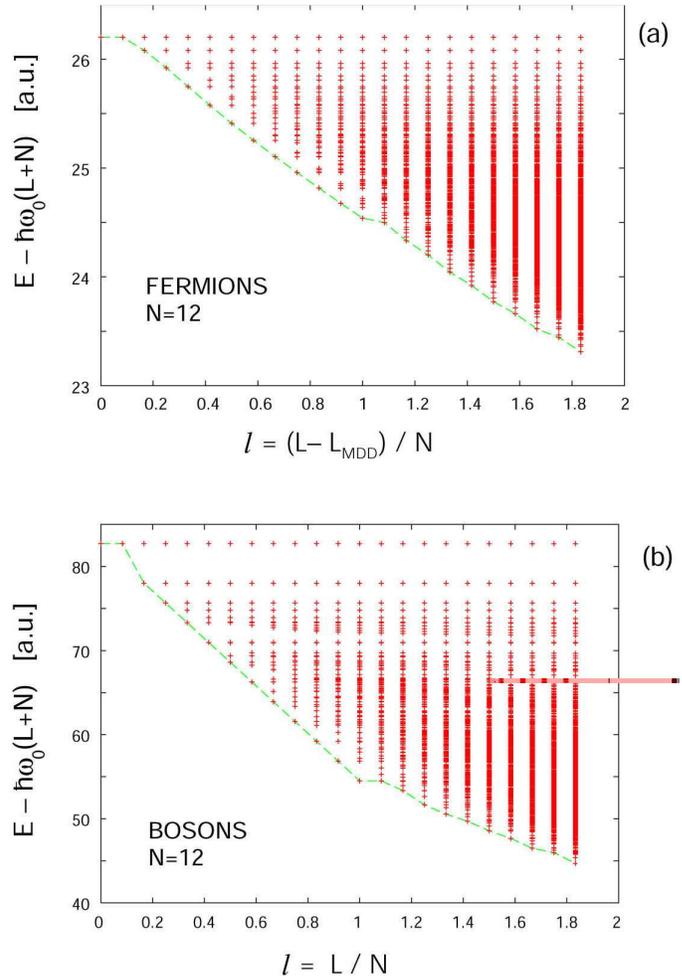}}
\caption{Interaction energy
versus angular momentum for a system of $12$ harmonically trapped, spin-less
particles interacting with a Coulomb repulsion, in (a) 
for fermions, and in (b) for bosons. The dashed line
is the so-called yrast line, connecting the lowest-energy states.}
\label{n12spec}
\end{figure}
\begin{figure}[!h]
\unitlength1cm
\centerline{\epsfxsize=9cm\epsfbox{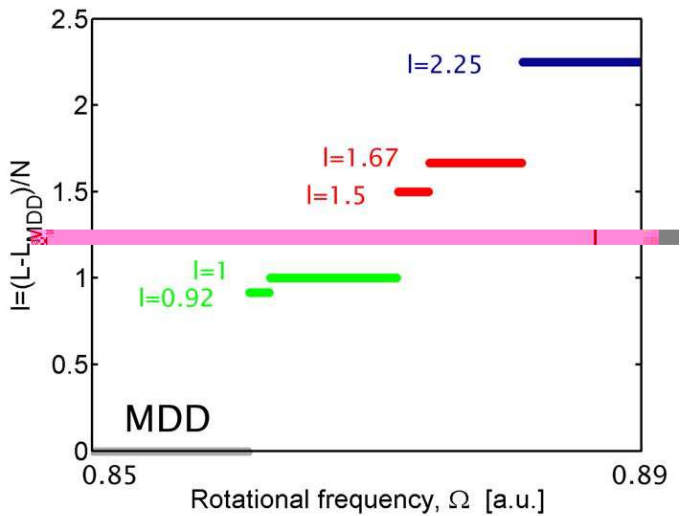}}
\caption{
  By taking $E_{L}-\Omega L$ for increasing total angular momentum the ground
  state values of $l=(L-L^{MDD})/N$ are found in a rotating
  system of 12 fermions as a function of the rotational frequency,
  $\Omega$. For increasing rotation the ground state is shifted from the MDD
  ($l=0$) to higher total angular momentum. In the figure successive
  transitions to states $l \approx 1$, $l \approx 1.6$ and to $l=2.25$
  occur. For computational reasons higher $l$-values have not considered.}
\label{n12omega}
\end{figure}

The interaction energy versus angular momentum is displayed in
Fig.~\ref{n12spec} for fermions (a) and bosons (b). Following the
tradition in nuclear physics, the dashed 
line connecting the lowest states
at fixed $L$ is called the ``yrast'' line \cite{yrast}, and the 
spectrum of $H$ the yrast spectrum. 
The interaction energy is plotted as a function of the above defined 
unit less variable $l$, which in a boson system corresponds to
$l=L/N$ and in a fermion system to $l=(L-L_{MDD})/N$. By this
construction we simplify the comparison between boson and fermion
systems, and it also becomes easier to compare 
systems of differing particle number. 
Note that $l=0$ in the fermion system corresponds to the MDD. In the boson
system, it describes the $L=0$ ground state.

Not surprisingly, both in the boson and fermion spectra 
(as shown for $N=12$ in Figure~\ref{n12spec}) the interaction energy
decreases with increasing $l$: A larger $l$-value implies a
higher rotation, which pushes the particles further apart.
However, the spectra  reveal other striking similarities. 
Recalling the fact that, due to the Pauli principle, 
the fermion total angular momenta are 
displaced with respect to the boson systems by the 
angular momentum of the MDD, we realize that the spectra
for fermions and bosons are very similar.
The yrast line in both cases shows a pronounced kink  at $l=1$.
and fainter kinks are observed 
at $l=1.67$ and at $l=1.83$ (not shown in Fig.~\ref{n12spec}).
For the boson system we know these particular $l$-values to be
related to the formation of vortices. 
According to the previously discussed results of the Gross-Pitaevskii 
equations a one-vortex solution corresponds to exactly $l=1$, and a two-vortex
solution to $l\approx 1.75$. What happens in the fermion 
system? 

Before studying the many-particle states in more detail, 
it is interesting to explore which $l$-values correspond to ground states 
in the rotating system. 
In Figure~\ref{n12omega} the ground states for increasing 
$\Omega$ have been found by subtracting $\Omega L$ from $E_L$.
Intriguingly, most of the $l$-values corresponding to 
kinks of the yrast line,  
including intermediate values $l=0.92$ and $1.5$, now reappear. 
(Note that to increase $\Omega$ above $1$~a.u. has no physical meaning. If the
rotational frequency exceeds the confining frequency the 
effective confinement disappears, see appendix.) 

To extract information from the
many-particle wavefunctions, one must process the data in some way. Two
informative quantities are the occupancies, $P_{m}$, of the single particle
states, which illustrates to what extent each single particle state (in the
LLL) is present in the many-body configuration, and the radial particle
densities,
\begin{displaymath}
\rho (\mathbf{r})=\langle \Psi_{F,B}(L) \vert \sum_{i=1}^{N} \delta(\mathbf{r}
- \mathbf{r_{i}}) \vert \Psi_{F,B}(L) \rangle,
\end{displaymath} 
where ${\bf r } = (r,0)$ in polar coordinates. 
The occupancies describe which single particle states 
dominate. If, for instance,  
$P_{m=0}$ has a magnitude close to zero the radial density
will be very small at $r=0$, since the single particle state (in the
LLL) with $m=0$ has its maximum at $r=0$. Similarly a
minimum at $P_{m=1}$ implies a slightly off-centred minimum in the
radial density. Hence, a central single vortex should appear as a minimum for
$m=0$ in the occupancy and at $r=0$ in the radial density, and a
two-vortex state as an off-centred minimum both in the occupancy and
in the radial density.
\begin{figure}[!h]
\unitlength1cm
\centerline{\epsfxsize=9cm\epsfbox{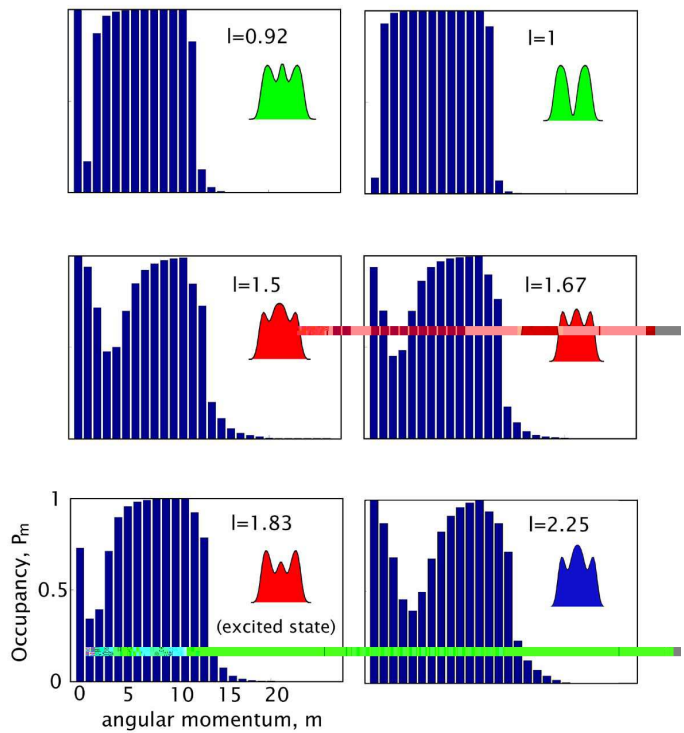}}
\caption{The occupancies of the single particle states for a system of
$N=12$ fermions. The values $l=0.92, 1, 1.5, 1.67$ and $2.25$ 
correspond to the ground state values
  in a system rotating at an increasing frequency. The excited state $l=1.83$
  has been included for completeness. The insets show the
  radial densities from $r=-6a_{B}^{*}$ to $r=6a_{B}^{*}$ for each
  $l$-value.}\label{n12occdens}
\end{figure}
In Figure~\ref{n12occdens} the occupancies are plotted for the 
ground state values $l=0.92, 1, 1.5, 1.67$ and $2.25$, and the excited state
$l=1.83$. The insets show the radial densities.
The last state has been included for completeness.
If we compare with the ground states in the bosonic system, these are
$l=1, 1.67, 1.83, $ and $2.25$. 
Even if the state $l=1.83$ in the fermion system 
is not a ground state in the rotating system, it is the lowest-energy state 
at that particular $l$ in the yrast spectrum. 
In the boson case we
know that a central single vortex appears at $l=1$ and a two-vortex solution
is found at $l
\approx 1.75$. As expected, in the fermion system, the central single
vortex is clearly visible at $l=1$. The two-vortex state,
however, is more difficult to identify. In contrast to mean field methods, for an exact calculation with good
angular momentum the particle density has circular symmetry and thus
does not display the internal structure directly. The vortices
are averaged over the entire circle and only show as two dips in the radial
densities. 
The ground state with $l=1.67$ and the excited state with 
$l=1.83$ showing two minima in the radial 
density thus constitute likely candidates for a two-vortex state. 

Another implication of many-vortex solutions 
in the exact result can be found in the
occupancies, which display some interesting systematics. To understand these
systematics better we chose to study only the structure of the Fock
state with the largest magnitude in the expansion  of $\vert
\Psi_{F}(L)\rangle$ (see Eq.~\ref{expansion}).
\begin{table}[!h]
\begin{center}
\begin{tabular}{c|c|c}
$l$ & $\vert c_{max} \vert^{2}$ & Many-body configuration \\
\hline
$0.92$ & 0.83 & 1011111111111000\\
$1.00$ & 0.92 & 0111111111111000\\
$1.50$ & 0.40 & 1110011111111100\\
$1.67$ & 0.44 & 1100111111111100\\
$1.83^*$ & 0.55 & 1001111111111100\\
$2.25$ & 0.28 & 1110011111111100\\
\end{tabular}
\caption{The dominating many-particle configurations for $N=12$ for the ground
state $l$-values. Note the single, respectively double, hole for $l=1$ and
$l=1.83$, which are the expected central single and double vortex
states. ($l=1.83$ is an excited state, as marked by a star).
}\label{tabell}
\end{center}
\end{table}

In Table~\ref{tabell} the
dominating configurations and their amplitudes are shown for
ground state $l$-values and the excited state $l=1.83$ 
in the case of $N=12$. For $l=0.92$ a single hole  has
entered the MDD. This corresponds to
the minimum shown in the occupancy in Figure~\ref{n12occdens}. At $l=1$ it
has reached the centre of the cloud and the single hole state has
become very dominant in the many-body configuration (consider its
magnitude). Correspondingly, the radial density shows a clear
central single vortex. At $l=1.5$ a double hole starts to enter the MDD. It
moves further towards the centre at $l=1.67$ and in the excited state 
$l=1.83$ it has reached the position, where its amplitude is the largest. This
configuration seems
to be a double hole close to the centre of the fermion cloud. Could it
be a two-vortex state? At $l=2.25$ a triple hole begins to enter the MDD.

As compared to mean-field approaches, these ``exact'' solutions 
retain the circular symmetry of the Hamiltonian. In order to reveal 
the internal structure of the many-body wave function, 
methods such as calculating the pair correlation, looking 
at a slightly asymmetric confinement potential~\cite{saarikoski2004b} or
adding a small perturbation~\cite{toreblad}
have to be used. 
Especially for small $N$, the above methods have shown useful: 
When, for example, studying pair correlations, 
fixing one of the very few particles was found to cause too  
much disturbance in the fermion many-particle configuration to get
a meaningful result.\\

\section{Pair Correlations}\label{pair}
The pair correlation fixes one particle at a point
$\mathbf{r_{A}}$ and 
calculates the density of the remaining $N-1$ particles.
\begin{displaymath}
  P(\mathbf{r},\mathbf{r_{A}}) =
  \frac{ \langle \Psi_{F,B}(L)\vert \sum_{i\ne
  j}\delta(\mathbf{r}-\mathbf{r_{i}}) 
  \delta (\mathbf{r_{A}} - \mathbf{r_{j}}) \vert \Psi_{F,B}(L) \rangle }
     {(N-1) \langle \Psi_{F,B}(L) \vert \sum_{j} \delta (\mathbf{r_{A}} -
       \mathbf{r_{j}}) \vert \Psi_{F,B}(L) \rangle }. 
\end{displaymath}

For large bosonic systems this shows the vortices clearly, see for instance
Kavoulakis \textit{et al} \cite{KavReiMot}. 

However, in the limit of small $N$, particularly in the fermion case,  
the removal of one particle effects the density
of the remaining particles significantly, and the result can be 
ambiguous:  Due to the Pauli principle the point, $\mathbf{r_{A}}$, at which 
one fermion has been fixed, will be
avoided by the other fermions. A minimum will arise in the pair
correlation at this point called the exchange hole. This makes the detection 
of vortices in the pair correlations very tedious. 
Particularly in the limit of small $N$, a small displacement 
of the reference point ${\bf r}_A$ alters the pair correlations 
considerably. 

\begin{figure}[!h]
\unitlength1cm
\centerline{\epsfxsize=7cm\epsfbox{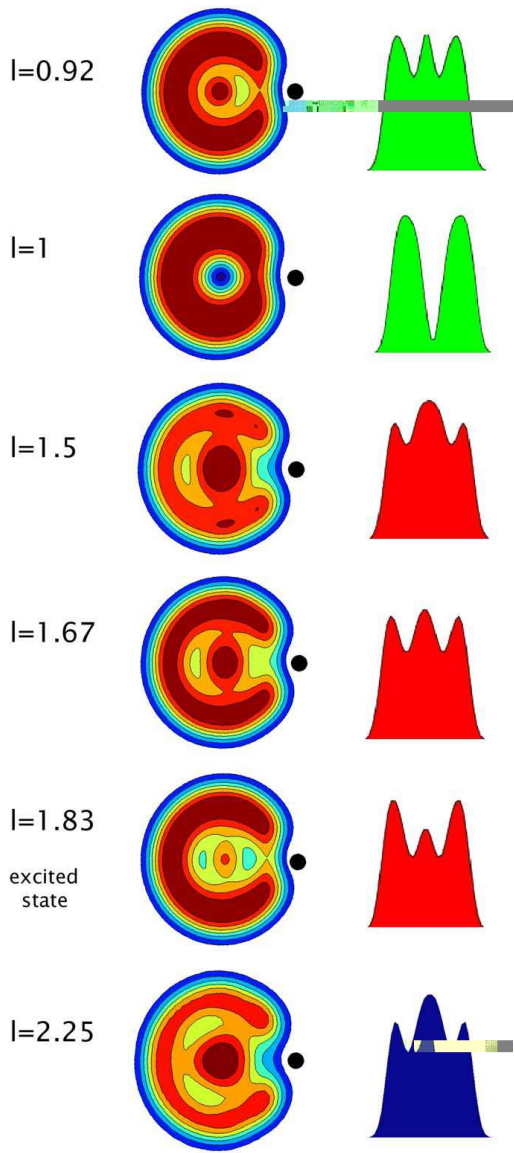}}
\caption{
Contours of the pair correlation {\it (left panel; 
the black dots mark the exchange hole
and the scale goes from minimum at dark blue to maximum at dark red)}
and radial density profiles ({\it right panel}) for 
$N=12$ at $l=0.92, 1, 1.5, 1.67, 1.83$ and $2.25$.
The vortices seem to enter the cloud from the outer regions, and then move
towards the centre. At $l=1$ a central single vortex can be seen and at
$l=1.83$ a two-vortex solution exists. }
\label{n12corrdens} 
\end{figure}

\begin{figure}[!h]
\unitlength1cm
\centerline{\epsfxsize=6cm\epsfbox{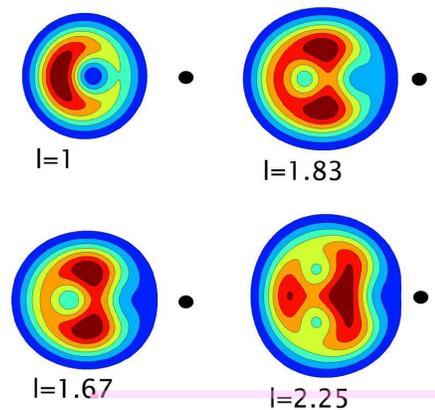}}
\caption{
Contours of the pair correlation for 
$N=12$ {\it bosons}, plotted as in Fig.~\ref{n12corrdens}. 
The reference point is marked with a black dot.}
\label{n12boscorr} 
\end{figure}

To come about this difficulty, 
two schemes have previously been suggested:  
$(i)$ adding a perturbation to the harmonic potential~\cite{toreblad}, 
and $(ii)$ breaking the rotational symmetry, for instance by 
deforming the confinement~\cite{saarikoski2004b}.
However, it appears that $N=12$ is sufficiently large to study the
pair correlations. The pair
correlations for $N=12$ have been plotted in Figure~\ref{n12corrdens} 
(fermions) and Figure~\ref{n12boscorr} (bosons) for
the ground state values of $l$ in both the fermion and bosonic case.
For completeness, the excited state $l=1.83$ has also been added in the fermion
case.
Apart from this state and the two intermediate states $l=0.92$ and $1.5$, 
there is an exact correspondence in the ground state 
$l$-values of the fermion and 
boson systems. With respect to angular momentum, this corresponds to the 
shift of the angular momentum $L_{MDD}$ 
of the maximum density droplet in the fermion system. 
Studying the pair
correlations for these $l$-values, the entering vortex in the rotating
system can clearly be seen, first changing to a central single vortex state at
$l=1$, and then to a double at $l=1.83$. 
This (excited) state is the same two-vortex solution as in the bosonic case, 
appearing at similar $l$. The solution at $l=2.25$ shows a third vortex
entering, as expected from the occupancies.

Note how the Pauli principle affects the fermion pair correlations 
by introducing the exchange hole (as marked by a black dot in 
Fig.~\ref{n12corrdens}). Naturally, this exchange hole is not present in the 
corresponding boson pair correlations. These are plotted in 
Figure~\ref{n12boscorr} for comparison. We note that although the pair
correlations in general are a bit ambiguous and the results depend 
strongly on the choice of the reference point, this method was 
earlier shown~\cite{KavReiMot} 
to yield a similar picture of vortex entry into the 
rotating cloud as the Gross-Pitaevskii approach~\cite{KavMotPet}. 

Compared to the mean-field picture the minima
will not equal zero due to the zero-point oscillations present in the
exact diagonalisation results.\\ 

\section{Edge reconstruction}

For particle numbers larger than approximately $15$ the central single vortex
disappears from the lowest state at $l=1$ and moves up into the spectrum of
excited states. 
It appears that, in the fermion system, the vortex configuration is not as
favourable anymore~\cite{yang}.

\begin{figure}[!h]
\unitlength1cm
\centerline{\epsfxsize=8cm\epsfbox{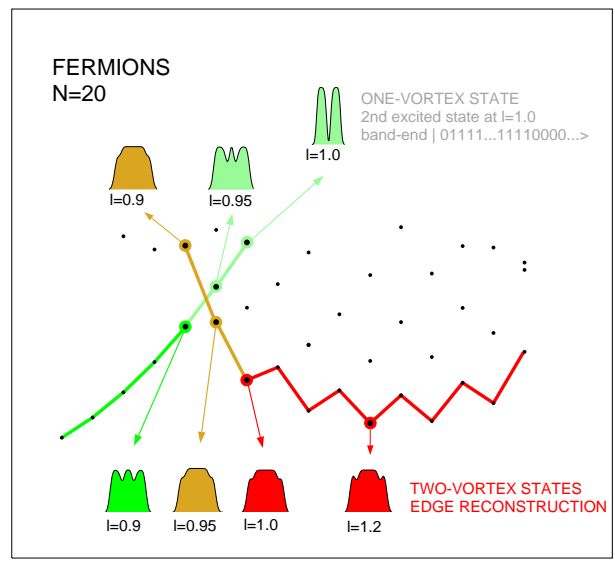}}
\caption{Low-lying many-body states as a function of $l$ for
  $N=20$ fermions. A polynomial $al^2 + bl+ c$ 
has been subtracted from the exact many-body energies to display the
  structure of the spectrum in greater detail.
Just below the central unit vortex, a level crossing occurs: The single vortex 
state (green line) moves upwards in energy, and is replaced by an 
edge reconstructed ground state.}\label{n20spec}
\end{figure}

\begin{figure}[!h]
\centerline{\epsfxsize=3.5in\epsfbox{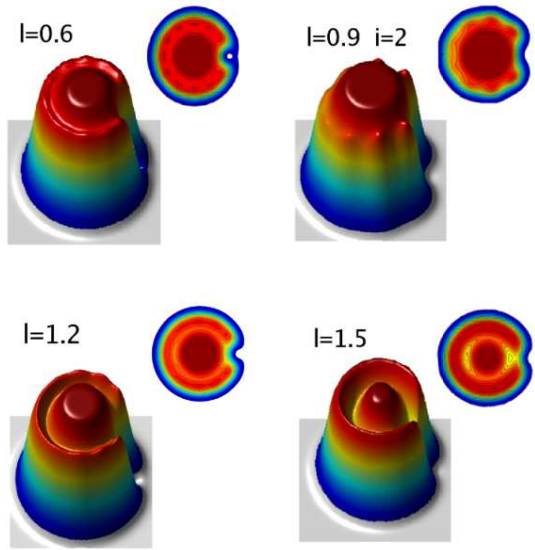}}
\caption{Pair correlations for N=20 fermions for some values of $l$ along the
  yrast line. In addition, $l=0.6$ and $l=1.2$ are ground states in the rotating
  system. Especially note the clear Chamon-Wen edge of the $l=0.6$ state,
  which can be compared to the CSDFT result in
  Reimann~{\it et al.}\cite{ReKoMa} with $l=0.75$.}\label{n20corr}
\end{figure}

In figure~\ref{n20spec} the yrast spectrum for $N=20$ fermions
 is displayed for increasing values of $l$. 
A second order polynomial, $al^2 + bl+c$, has been 
subtracted from the interaction
energy in order to display the structure of the spectrum in greater detail. 
In fact, below $l=0.9$, the entering vortex state dominates as it did 
for smaller particle numbers. However, at $l=0.9$, there exists a level
crossing in the yrast spectrum. The state with the entering vortex becomes an 
excited state, which corresponds to the central one-vortex configuration at 
$l=1$, while a state with a distinct edge 
becomes most prominent above $l=0.9$. 
Similarly, the $l$-values at which the actual ground states occur 
in the rotating spectrum, are altered. 
The central one-vortex, respectively the entering two-vortex 
at $l=1$ and $l\approx 1.7$, which we saw for 
$12$ fermions, no longer appear as ground states. Instead, configurations
 around $l=0.6$, $l=1.2$ and $l=1.8$ have become more favourable
in energy. All three constitute states with a clear edge, and in
particular $l=0.6$ and $l=1.2$ are interesting since their edges
consist of localised fermions, so-called Chamon-Wen edges
(see Figure~\ref{n20corr}).
This is in contrast to the rotating boson system 
which is known~\cite{butts,KavMotPet} 
to have a central one-vortex ground state at $l=1$.
\begin{figure}[!h]
\unitlength1cm
\centerline{\epsfxsize=9cm\epsfbox{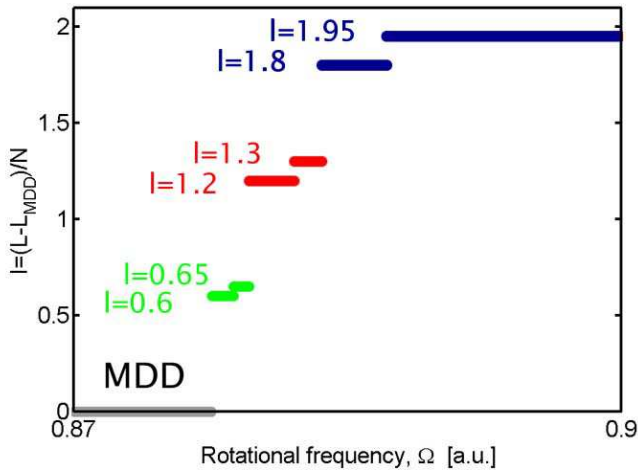}}
\caption{Ground state $l$-values for increasing rotation ($\Omega$) no longer
  contains the central single and double vortex states in the 20 fermion
  system. Instead edge reconstructed states, like $l=0.6$ and $l=1.2$ become
  dominating.}\label{n20omega}
\end{figure}

\begin{figure}[!h]
\unitlength1cm
\centerline{\epsfxsize=9cm\epsfbox{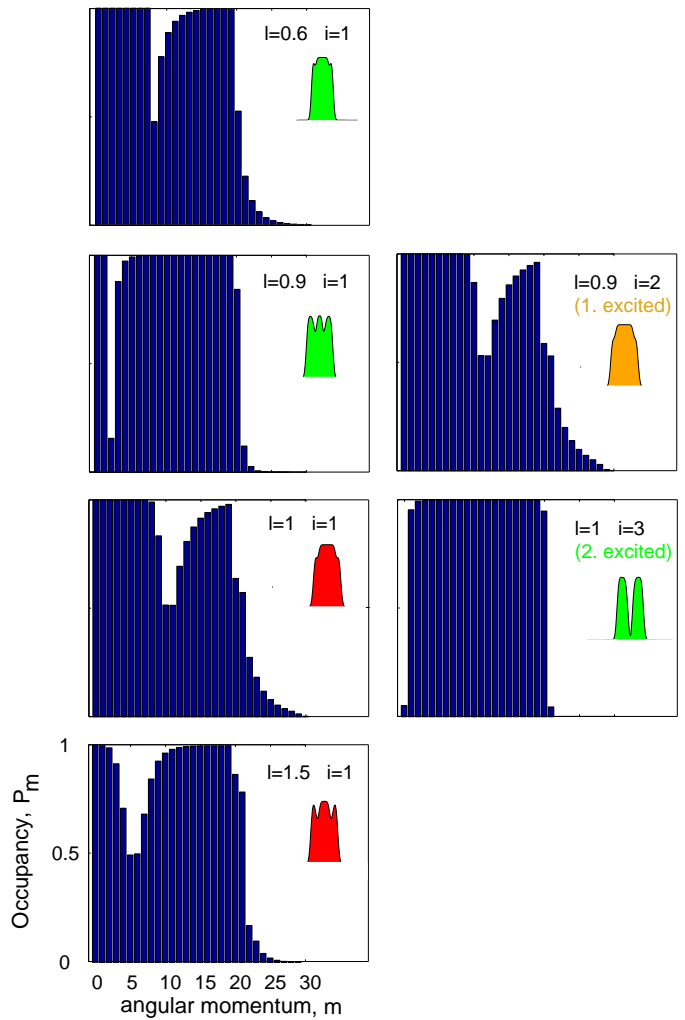}}
\caption{Occupancies of the single particle states for a system of
$N=20$ fermions, shown for the lowest states for increasing $l$. 
Corresponding
  radial densities from $r=-7a_{B}^{*}$ to $r=7a_{B}^{*}$ are shown in the insets. The
  central single vortex state has now disappeared from the lowest state
  $(i=1)$ and has become the 2nd excited state $(i=3)$ at
  $l=1$.}\label{n20occdens}
\end{figure}

A Chamon-Wen edge becomes favourable when a higher rotation (or magnetic
field)
effectively decreases the radius of the single particle states as compared to
the confining trap. For a
rotating system at high angular momentum this would
imply an increase in density relative to the confinement with increasing
rotational motion, hence
effectively compressing the system. The repulsive Coulomb interaction
strongly inhibits this, and the maximum density droplet in effect redistributes
its particles from lower to higher single particle momentum, leaving
some states with $m < N-1$ unoccupied, giving a ground state with $L
> N(N-1)/2$ \cite{ReKoMa}. Figure~\ref{n20occdens} shows the occupancies
and radial densities of some of the lowest energy states along the yrast line
for increasing $l$.
Both the occupancies and the radial densities imply a
Chamon-Wen edge, i.e. that there exists an edge of redistributed fermions
and an inner dense MDD droplet. 
Also visible is the entering single vortex configuration at $l=0.9$ 
which becomes the second excited state in the yrast spectrum at 
$l=1~(i=3)$. This is the central one-vortex state.

Studying also the pair correlations for $N=20$ pictured in
Figure~\ref{n20corr} for $l=0.6$ respectively $l=1.2$, the edge of localised
fermions is clearly visible. 
These pair correlations compare well with the results by 
Reimann \textit{et al.}~\cite{ReKoMa}, showing electron localisation 
along the Chamon-Wen edge within mean field current spin density functional
theory (CSDFT). 
Note that in the exact result the localisation
becomes less pronounced due to the zero point oscillations.
At the first glance one might draw the conclusion that the central single
vortex
disappears from the lowest energy at $l=1$ since we have increased the
number of fermions in the trap. This suggests a higher particle
density, which does not allow for a central minimum, the single
vortex. However, the interaction energy only scales with a factor depending on
$\omega_{0}$ (see Sec.~II). 
Hence, our results show that the central single vortex state
will no longer be
favourable irrespective of confinement frequency. 
\begin{figure}[!h]
\centerline{\epsfxsize=3.5in\epsfbox{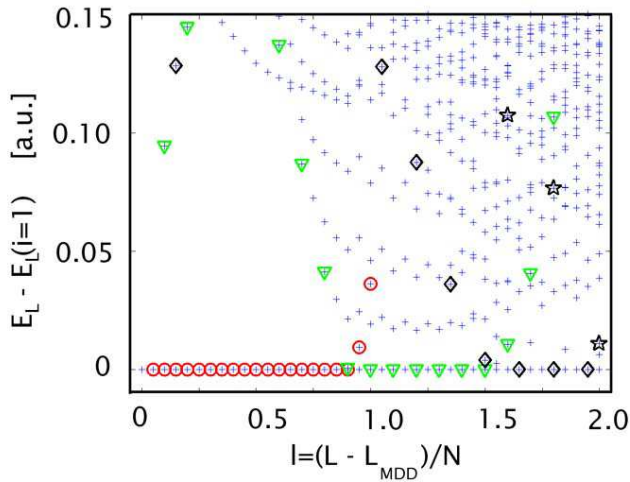}}
\caption{Yrast spectrum for $N=20$, with the lowest energy at fixed $l$
  ($E_{L}(i=1)$) subtracted. The (red) circles mark a single zero entering the
  Fock state of the MDD, the (green) triangles a double hole, for example
  $\mid 11\cdots 11100111\cdots 11000\cdots \rangle$  and (black) diamonds a
  triple hole. }\label{n20bands}
\end{figure}

Solutions where vortices enter at the edge of the droplet, 
can co-exist with edge reconstructed configurations.
The entering second vortex, as well as the (now off-centred) 
first vortex appear as two clear minima. Consider the pair correlation for
$l=1.2$ in Figure~\ref{n20corr}.
A more pronounced entering two-vortex state shows in the pair correlation of
$l=1.5$ along the yrast line, also in Figure~\ref{n20corr}. This agrees well
with the double hole in the occupancy of Figure~\ref{n20occdens} for the same
state.
The entering second vortex can also be seen in the energy spectrum in
Fig.~\ref{n20spec} as oscillations with a period of $\Delta l=0.1$. 
As explained earlier, by subtracting a second order polynomial, 
the oscillations in the yrast spectrum are enhanced. 
In Manninen{\it et al.}~\cite{holes} this was discussed in greater 
detail. It was found that the oscillations change period 
at higher $l$, first to become three-fold and then to become four-fold. 
In Figure~\ref{n20bands} we compare the oscillation periods with 
holes in the Fock states, by studying the dominating 
configurations for increasing $l$.
For simplicity, the lowest energy at each $l$-value
$(E_L (i=1))$ have been subtracted, i.e. we study the 
excitation spectrum. The (red) circles describe the entering vortex state, 
$| 11\cdots 11011\cdots 10\cdots \rangle $, which is seen to become the 
second excited state in the yrast spectrum at $l=1$. Earlier, this was 
found to be the central single vortex configuration. 
The (green) triangles are then seen to move downwards in the spectrum for
increasing $l$. These are the entering double vortex states,
$| 11\cdots 110011\cdots 10\cdots \rangle $, 
appearing in the energy spectrum as oscillations with period $\Delta l=0.1$. 
Moving the double hole (00) one step to the left implies an increase in total 
angular momentum of two. 
The entering double vortex, thus, only appears at every second $L$
(in the Figure $\Delta l=0.1$), corresponding to the double oscillations. 
Similarly, the entering triple vortex, (black) diamonds, with Fock state 
$| 11\cdots 1100011\cdots 10\cdots \rangle $, can only occur at every third
$L$, corresponding to a three-fold oscillation period. This is in exact
agreement with Manninen~{\it et al.}~\cite{holes}.

\section{Summary and conclusions}\label{summary}

To conclude, the results presented have shown that there indeed exists some
universal features between rotating systems of bosons and (unpaired)
fermions. 
All calculations have been performed using the method of exact
diagonalisation, and the wavefunction data have been evaluated with the aid of 
occupancies and pair correlations. 

Especially, the tendency of these systems to form
vortices at similar angular momenta have been explored. However, for
larger particle numbers than
approximately $15$ the central single vortex state becomes an excited state and
the ground state, instead,
lowers its energy by reconstructing to form an
edge of localised fermions. This is in agreement with earlier
mean-field computations, see for instance Reimann \textit{et
  el.}~\cite{ReKoMa}.

\section*{Acknowledgements}
We thank B. Mottelson, D. Pfannkuche, 
H. Saarikoski, E. R\"as\"anen, A. Harju and M. Puska 
for rewarding discussions. 
Financial support from the Swedish Foundation for Strategic Research
and the Swedish Research Council is gratefully acknowledged.
\\

{\bf APPENDIX: ANALOGY BETWEEN ROTATION AND MAGNETIC FIELD}

\medskip

Consider the Hamiltonian of the rotating system, Eq.~(\ref{hamiltonian}), 
in Sect.~II.
Re-writing $H_{\Omega }$, we show that it equally well corresponds 
to a harmonic oscillator
subject to an effective magnetic field of strength $B=\nabla \times
(m\Omega \hat{z}\times \bar{r}/e) = (2m \Omega /e) \hat{z}$ and with a
confinement frequency $(\omega_{0}^{2} - \Omega^{2})$~\cite{CooWilGun}:
\begin{eqnarray}
H_{\Omega} &
= & \sum_{i=1}^{N} \left(\frac{( \mathbf{p}_{i}-m\Omega \hat{z} \times
  \mathbf{r}_{i})^{2}}{2m}+ \frac{1}{2} 
m (\omega_{0}^{2}-\Omega^{2}) \mathbf{r}^{2}_{i}\right) \nonumber \\ 
& & + \sum_{i <j}^{N} \frac{e^{2}}{4 \pi \epsilon \epsilon_{0} |
  \mathbf{r}_{i} - \mathbf{r}_{j} |} \nonumber 
\end{eqnarray}
As a consequence, the rotation effectively lowers the confinement in
x-y-direction to $(\omega_{0}^{2} - \Omega^{2})$.

\end{document}